\documentstyle[12pt]{article}

\title{\bf $N$ electrons in a quantum dot:\\
           Two-point Pade approximants}

\author{Augusto Gonzalez$^*$}

\date{Depto. de Fisica, Univ. Nacional de Colombia, Sede Medellin,\\
       AA 3840, Medellin, Colombia}

\begin{document}

\maketitle

\begin{abstract}
We present analytic estimates for the energy levels of $N$ electrons
($N = 2 - 5$) in a two-dimensional parabolic
quantum dot. A magnetic field is applied perpendicularly to the confinement
plane. The relevant scaled energy is shown to be a smooth function of the
parameter $\beta$=(effective Rydberg/effective dot energy)$^{1/6}$.
Two-point Pade approximants are obtained from the series expansions of the
energy near the oscillator ($\beta\to 0$) and Wigner ($\beta\to\infty$)
limits. The approximants are expected to work with an error not greater
than 2.5\% in the
entire interval $0\le\beta<\infty$.

\end{abstract}
\vskip 5cm

$^*$ On leave from the Instituto de Cibernetica, Matematica y Fisica,
     Calle E 309, Vedado, Habana 4, Cuba

\section{Introduction}

The problem of $N$ electrons in quantum dots and magnetic fields has been
widely considered in recent years [1].

Concerning the energy levels of $N$ electrons in model two-dimensional
parabolic quantum dots, the actual magnitude to compute is a scaled energy
which depends only on one parameter, $\beta$=(effective Rydberg/effective
dot energy)$^{1/6}$. When $\beta\to 0$ (very high magnetic fields, for
example), we may use perturbation theory to compute the energy levels. In
the opposite limit, $\beta\to\infty$, a strong coupling expansion may be
used to obtain the energy. The idea of the present paper is to construct
two-point Pade approximants interpolating from $\beta=0$ to
$\beta\to\infty$. We show results for 2, 3, 4, and 5 electrons.

To our knowledge, there are only a few alternative analytical ways to
obtain the energy of certain levels in the entire interval
$0\le\beta<\infty$. Semiclassical [2] and 1/$|J|$-expansions [3, 4], both
working for states with high angular momentum, $J$, are available.
Besides these expansions, there is also an idea of improving the
perturbative series by using the asymptotics at $\beta\to\infty$ [5]. We
understand the present paper as a useful complement to the results of
[2 - 5]. We will see, for example, that it is very simple to find Pade
approximants for states with $|J|$= 0, 1, 2, for which the methods of [2, 4]
do not work.

\section{Two-point Pade approximants}

The construction of approximants follows the idea of paper [6], in which
the hydrogenic energy levels in a magnetic field were obtained.

Let us consider the expansions of the scaled energy (to be computed in the
next sections) when $\beta\to 0$ and $\beta\to\infty$,

\begin{eqnarray}
\left. \epsilon\right|_{\beta\to 0} &=& \sum_{k=0}^{s} b_k \beta^k
                                       + {\cal O}(\beta^{s+1}),\\
\left. \epsilon\right|_{\beta\to\infty} &=& \beta^2 \left\{ \sum_{k=0}^{t}
                             a_k/\beta^k + {\cal O}(1/\beta^{t+1})\right\}.
\end{eqnarray}

\noindent
Many of the coefficients entering (1, 2) are zero. For example, all the
$b_k$ with $k\ne 0~~mod~3$, $a_1$, $a_3$, etc.

A two-point Pade approximant is a rational function,

\begin{equation}
P(\beta) = {\sum_{k=0}^L p_k \beta^k \over \sum_{k=0}^K q_k \beta^k},
\end{equation}

\noindent
reproducing the expansions (1) and (2). $q_0$ may be fixed to one. The
asymptotics when $\beta\to\infty$ forces $L$ to be equal to $K+2$.
Equating number of coefficients in the expansions (1) and (2) to number of
unknowns in (3), we obtain,

\begin{equation}
s + t = 2 K + 1.
\end{equation}

At a given $K$, there is a set of possible pairs $s, t$. We will attach
indexes $s$ and $t$ to the approximant, $P_{s,t}$. Among the possible
$P_{s,t}$, the best one takes nearly the same number of terms in the
expansions (1) and (2) [6], i.e. $s \approx t$. For example, at $K=3$ the
best one, which coefficients are computed almost trivially, is $P_{4,3}$.

Let us consider the equations fulfilled by the $p_k$ and the $q_k$.
Equating (3) to (1) and (2), we obtain,

\begin{eqnarray}
p_k &=& q_0 b_k +q_1 b_{k-1}+\dots+q_k b_0,~~~0\le k\le{\rm Min}(s, K+2), \\
p_k &=& q_K a_{K+2-k} +q_{K-1} a_{K+1-k}+\dots+q_{k-2} a_0,\nonumber\\
     && K+2-{\rm Min}(K+2,t)\le k\le K+2.
\end{eqnarray}

\noindent
Eliminating the overlapping $p_k$, i.e. $k$ in the interval
${\rm Max}(0, s+1-K)\le k\le{\rm Min}(s, K+2)$, a system of linear
equations for the coefficients $q_k$ arise. We assume that $s$ is in the
interval $K-1\le s\le K+2$, such that this system contains $K$ equations.
Let us write explicitly, for
example, the approximant $P_{4,3}$. The coefficients $b_1$, $b_2$, $b_4$,
$a_1$ and $a_3$ are
assumed to be zero. The result is the following,

\begin{eqnarray}
P_{4,3}(\beta)=b_0+\frac{b_3\beta^3}{1+q_1\beta+q_2\beta^2+q_3\beta^3}
 + a_0\beta^2 \left\{1-\frac{1+q_1\beta}{1+q_1\beta+q_2\beta^2+q_3\beta^3}
 \right\}, \nonumber \\
{\rm where}~q_2=a_0/(b_0-a_2),~q_1=a_0 q_2/b_3,~q_3=(a_0 q_1-b_3)/(b_0-a_2).
\end{eqnarray}

Formula (7), or similar expressions for higher approximants (see Appendix
1), is to be used throughout the paper. Let us indicate the way to compute
the $b_k$ and the $a_k$.

\section{The coefficients $b_k$ and $a_k$}

Let us consider the two dimensional motion of $N$ electrons in a parabolic
quantum dot of energy $\hbar\omega_0$. A magnetic field is applied normaly
to the plane of motion. The hamiltonian governing the internal motion
(center of mass motion is excluded) is written in dimensionless variables as

\begin{equation}
\frac{H}{\hbar\Omega} = h + \frac{\omega_c}{2\Omega} J + \frac{g\omega_c}
                             {2\Omega}S_z,
\end{equation}

\noindent
where $\omega_c$ is the cyclotronic frequency, $\Omega=\sqrt{\omega_0^2+
\omega_c^2/4}$ is the effective dot frequency, $J$ is the total (internal)
angular momentum (along the $z$ axis), $S_z$ is the $z$-projection of the
total spin, $g$ is the effective giromagnetic factor, and

\begin{eqnarray}
h &=& -\sum_{k=1}^{N-1} \left(
          \frac{\partial^2}{\partial\rho_k^2} + \frac{1}{\rho_k}
          \frac{\partial}{\partial\rho_k} \right)
        - \sum_{k=1}^{N-2} \left( \frac{1}{\rho_k^2}+\frac{1}{\rho_{k+1}^2}
          \right) \frac{\partial^2}{\partial\theta_k^2} \nonumber \\
    &+& 2 \sum_{k=1}^{N-3} \frac{1}{\rho_{k+1}^2}
          \frac{\partial^2}{\partial\theta_k\partial\theta_{k+1}}
        + \frac{2 i J}{N-1} \sum_{k=1}^{N-2} \left( \frac{1}{\rho_k^2}
          - \frac {1}{\rho_{k+1}^2} \right)\frac{\partial}{\partial\theta_k}
          \nonumber \\
    &+& \sum_{k=1}^{N-1} \left( \frac{J^2}{(N-1)^2} \frac{1}{\rho_k^2}
          + \frac{1}{4} \rho_k^2 \right)
        + \beta^3 \sum_{k<l} \frac{1}{r_{kl}}
\end{eqnarray}

\noindent
Notice that $h$ depends only on one parameter, $\beta^3=\sqrt{\mu e^4/
\kappa^2 \hbar^2}/\sqrt{\hbar\Omega}$, where $\mu$ is the effective
electron mass, and $\kappa$ - the relative dielectric constant. The
coordinates entering $h$ are the moduli of the Jacobi vectors,

\begin{equation}
\vec\rho_k = \sqrt{\mu_k\over\mu_1} \left\{\vec r_{k+1} - \frac{1}{k}
             \sum_{j=1}^{k} \vec r_j \right\},~~k=1,\dots,N-1,
\end{equation}

\noindent
and the angles between
$\vec\rho_k$ and $\vec\rho_{k+1}$, denoted by $\theta_k$. The dimensionless
reduced masses are $\mu_k=k/(k+1)$.

The eigenfunctions of $H$ are $e^{i J \Xi} \psi$, where $\Xi$ accounts for
global rotations, and $\psi$ are the eigenfunctions of $h$. The eigenvalues
of $H$ are trivially obtained from the eigenvalues of $h$, which will be
called $\epsilon$. We will obtain Pade approximants to $\epsilon$.

In the $\beta\to 0$ (oscillator) limit, perturbation theory may be applied
to obtain $\epsilon$. The resulting series is the following

\begin{equation}
\left. \epsilon\right|_{\beta\to 0} = b_0+b_3\beta^3+b_6\beta^6+\dots,
\end{equation}

\noindent
where $b_0=N-1+|J|+2 n$, $n$ is the total number of oscillator quanta,
$b_3=\left< \phi|\sum_{k<l} r_{kl}^{-1} |\phi\right>$, etc. In systems with
more than two electrons, sometimes degenerate perturbation theory shall be
used to compute $b_3$, $b_6$, etc.

On the other hand, when $\beta\to\infty$ a strong coupling expansion may
be applied. Distances are scaled according to $\rho\to\beta\rho$. The
hamiltonian $h$ takes the form

\begin{eqnarray}
h\over\beta^2 &=& \frac{1}{4} \sum_{k=1}^{N-1} \rho_k^2
        + \sum_{k<l} \frac{1}{r_{kl}}
        + \frac{J^2}{(N-1)^2 \beta^4} \sum_{k=1}^{N-1} \frac{1}{\rho_k^2}
        \nonumber \\
    &-& \frac{1}{\beta^4} \left\{ \sum_{k=1}^{N-1} \left(
          \frac{\partial^2}{\partial\rho_k^2} + \frac{1}{\rho_k}
          \frac{\partial}{\partial\rho_k} \right)
        + \sum_{k=1}^{N-2} \left( \frac{1}{\rho_k^2}+\frac{1}{\rho_{k+1}^2}
          \right) \frac{\partial^2}{\partial\theta_k^2}
          \right \} \nonumber \\
    &+& \frac{2}{\beta^4} \left\{ \sum_{k=1}^{N-3} \frac{1}{\rho_{k+1}^2}
          \frac{\partial^2}{\partial\theta_k\partial\theta_{k+1}}
        + \frac{i J}{N-1} \sum_{k=1}^{N-2} \left( \frac{1}{\rho_k^2}
          - \frac {1}{\rho_{k+1}^2} \right)\frac{\partial}{\partial\theta_k}
          \right\}.
\end{eqnarray}

In the leading approximation, $\beta\to\infty$, we shall minimize the
classical potential energy entering the r.h.s. of (12). It is found that
the electrons sit at the corners of a
regular polygon. This is a few-body version of the Wigner solid. The
classical energy becomes a function of one variable, $\rho_1$,

\begin{equation}
U = \frac{N \rho_1^2}{8 \sin^2 \pi/N} +
    \frac{\sin \pi/N}{\rho_1} \sum \frac{1}{|\sin\theta_{kl}/2|},
\end{equation}

\noindent
where $\theta_{kl}$ is the angle between particles $k$ and $l$, measured
from the c.m. The minimization of $U$ leads to the equilibrium value of
$\rho_1$, $\rho_{10}$. The equilibrium values of the other coordinates are
obtained from geometric considerations. The energy in this approximation
is given by $U(\rho_{10})$. Corrections to this value are obtained by
writting $\rho_k=\rho_{k0}+z_k/\beta$, $k=1,\dots,N-1$, $\theta_k=
\theta_{k0}+z_{N-1+k}/\beta$, $k=1,\dots,N-2$, and expanding

\begin{eqnarray}
h/\beta^2 &=& U(\rho_{10})+h_2/\beta^2+h_3/\beta^3+\dots,\\
\psi &=& \psi_0+\psi_1/\beta+\psi_2/\beta^2+\dots,\\
\epsilon/\beta^2 &=& a_0+a_2/\beta^2+a_3/\beta^3+\dots.
\end{eqnarray}

The Schrodinger equation is split into the set of uncoupled equations

\begin{eqnarray}
a_0 &=& U(\rho_{10}),\\
h_2 \psi_0 &=& a_2 \psi_0,~~{\rm etc}.
\end{eqnarray}

\noindent
The hamiltonian $h_2$ describes harmonic oscillations around the Wigner
structure. The expression for $a_2$ is thus

\begin{equation}
a_2 = \sum_{k=1}^{2 N-3} \omega_k (n_k+1/2),
\end{equation}

\noindent
where the $\omega_k$ are the normal frequencies. Higher corrections are
obtained by considering $h_3$, $h_4$, etc as perturbations to $h_2$.

Below, we present results for 2, 3, 4, and 5 electrons.

\section{Two electrons}

As mentioned above, the eigenfunctions of $H$ are written as $e^{i J \Xi}
\psi(\rho_1)$. Under a permutation of particles, $\Xi$ changes by $\pi$,
and $\psi$ does not change. Thus, even $|J|$ are related to antisymmetric
spin functions, $S=0$, and odd $|J|$ are related to states with spin
$S=1$.

Let $\phi_k$ be the eigenfunctions of $h$ at $\beta=0$,

\begin{equation}
\phi_k = C_k \rho_1^{|J|} L_k^{|J|}(\rho_1^2/2) e^{-\rho_1^2/4},
\end{equation}

\noindent
where $C_k=(2^{|J|}(k+|J|)!/k!)^{-1/2}$, and $L_k^{|J|}$ are
generalized Laguerre polynomials. The corresponding eigenvalues
are

\begin{equation}
b_0 = 2 k + |J| + 1.
\end{equation}

\noindent
We take a fixed $k$ as the unperturbed level, let us say, $n_1$.
Higher coefficients of the expansion are computed from

\begin{eqnarray}
b_3 &=& <n_1|\frac{1}{\rho_1}|n_1>,\\
b_6 &=& {1 \over 2} \sum_{k\ne n_1}{|<n_1|1/\rho_1|k>|^2 \over (n_1-k)},
        ~~{\rm etc}.
\end{eqnarray}

We show in Tab. 1 the coefficients $b_3$ and $b_6$ for a set of
two-electron states.

Let us consider now the opposite limit, $\beta\to\infty$. The equilibrium
value of $\rho_1$ (scaled) is $\rho_{10}=2^{1/3}$. The coefficient $a_0$
is thus $a_0=3/2^{4/3}$. Then, we write $\rho_1=\rho_{10}+z_1/\beta$, and
expand the hamiltonian. The results are

\begin{eqnarray}
h_2 &=& -{\partial^2 \over \partial z_1^2}+ {3\over 4} z_1^2,\\
h_k &=& {(-1)^k \over \rho_{10}^{k+1}} z_1^k -
   {(-1)^{k-3} \over \rho_{10}^{k-2}} z_1^{k-3}{\partial\over\partial z_1}+
   {J^2 (-1)^{k-4}(k-3) \over \rho_{10}^{k-2}} z_1^{k-4},~~k\ge 3.
\end{eqnarray}

Notice that the $h_k$ with odd $k$ contain an odd number of creation and
anihilation operators. Thus, all the $a_k$ with odd $k$ will be zero.
Computation of the matrix elements of $h_k$ is a trivial task. Finally,
we obtain

\begin{eqnarray}
a_2 &=& \omega_1 (n_1 + 1/2),\\
a_4 &=& 2^{-2/3} \left\{ (n_1^2+n_1+7/6)/6 + J^2 -1/4 \right\},~~{\rm etc}
\end{eqnarray}

\noindent
where $\omega_1=\sqrt{3}$, i.e. the classical result [7]. Note that we
have used the same number, $n_1$, to label the state at $\beta=0$ and
$\beta\to\infty$. As level crossings can not occur, the first state at
$\beta=0$ should be the first when $\beta\to\infty$, and so on. Note
also that $J$ appears for the first time in $a_4$.

The coefficients $a_2$ and $a_4$ are also shown in Tab. 1 for a set of
states. From this Table and formulae contained in Appendix 1, we may
construct Pade approximants for the energy levels.

We show in Fig. 1 the three curves $\left.\epsilon\right|_{\beta\to 0}=
b_0+b_3 \beta^3+b_6 \beta^6$, $\left.\epsilon\right|_{\beta\to\infty}=
a_0 \beta^2+a_2 + a_4/\beta^2$, and $P_{6,5}(\beta)$ for the first state
with quantum numbers $|J|=3$, $n_1=0$. This is the typical behaviour of the
approximants.

In Fig. 2, we compare the approximants in the sequence $P_{K+1,K}$. The
same state as in Fig. 1 is studied. The relative differences between
consecutive approximants are shown. The $\beta$ axis is compressed to
(0,1). We see that the maximal difference reduces by a factor of two when
$K$ is increased by one. These results suggest the $P_{6,5}$ approximant
(the highest we computed) to be accurate to about three parts in $10^3$
or better. Notice that the maxima are reached at $\beta\approx 2$, i.e.
in the region where the approximants jump from the weak-coupling to the
strong-coupling regimes (see Fig. 1).

Similar results are obtained for the other states in Tab. 1. When $J=0$,
however, relative differences between approximants increase up to 2\%,
and some approximants can not be used as they exhibit a pole. So, the best
of our Pade estimates are expected to work with an error not greater than
2\% in the worst situation.

We show in Fig. 3 the convergence of the sequence $P_{K+1,K}$ at particular
$\beta$ values, at which exact solutions are available [8]. It can be easily
shown, for example, that

\begin{equation}
\psi = \rho_1^{|J|} \left( 1 + \rho_1/\sqrt{2 |J|+1} \right) e^{-\rho_1^2},
\end{equation}

\noindent
are eigenfunctions of $h$ at $\beta= (2 |J|+1)^{1/6}$, with eigenvalues
$\epsilon=|J|+2$. The corresponding $n_1$ are zero. When $0\le |J|\le 3$,
$\beta$ is in the interval $1\le \beta\le 1.38$, well outside the exactly
solvable limits. The relative error of the $P_{6,5}$ approximant is lower
than 2 \% at $J=0$, and less than 0.2 \% at $|J|=3$.

\section{Three electrons}

First, let us consider the computation of the coefficients $a_k$ up to
$a_5$. The Wigner configuration is a triangle with side $\rho_{10}=3^{1/3}$
(scaled). The leading approximation to the energy is

\begin{equation}
a_0 = {3^{5/3} \over 2}.
\end{equation}

Then, we write $\rho_1=\rho_{10}+(u+v)/(\sqrt{2}\beta)$,
$\rho_2=\rho_{10}+(u-v)/(\sqrt{2}\beta)$,
$\theta_1=\pi/2+\sqrt{2}z/(\rho_{10}\beta)$. Expanding the hamiltonian $h$,
we arrive to

\begin{eqnarray}
h_2 &=&-\left( {\partial^2 \over \partial u^2}+
               {\partial^2 \over \partial v^2}+
               {\partial^2 \over \partial z^2}\right)
       + {3\over 4} u^2 + {3\over 8} (v^2+z^2)     ,\\
h_3 &=& {1 \over 16\sqrt{2}\rho_{10}} \left\{
        -32{\partial\over\partial u}+32 u {\partial^2 \over\partial z^2}
        -8 u^3-6 u v^2-5 v^3-6 u z^2 \right. \nonumber\\
    & &\left. +15 v z^2 \right\}, \\
h_4 &=& {1 \over \rho_{10}^2} \left\{ {J^2 \over 2}
  +u{\partial\over\partial u}-{3\over 2} u^2 {\partial^2 \over\partial z^2}
  +{1\over 4} u^4 + v{\partial\over \partial v}
  -2 i J v{\partial\over \partial z} \right. \nonumber\\
    & &-{3\over 2} v^2{\partial^2\over \partial z^2} +{3\over 8}u^2 v^2
  +{5\over 8}u v^3 +{9\over 256}v^4 +{3\over 16}u^2 z^2 \nonumber \\
    & &\left. -{15\over 16}u v z^2 -{3\over 128}v^2 z^2 +{41\over 256}z^4
       \right\},~~{\rm etc}.
\end{eqnarray}

$u$, $v$, and $z$ are normal coordinates. $u$ corresponds to a symmetric
oscillation (breathing) with frequency $\omega_1=\sqrt{3}$, whereas $v$
and $z$ correspond to a mixed oscillation of the Wigner structure, with
frequency $\omega_2=\sqrt{3/2}$. In the harmonic approximation, the spatial
wave function is written as

\begin{equation}
\psi = \chi_{n_1}(u) \chi_{n_2}(v) \chi_{n_3}(z) e^{i J \Xi},
\end{equation}

\noindent
where the $\chi$ are oscillator functions, i.e. Hermite polynomials
multiplied by gaussians. A function $\psi$ that can be antisymmetrized is
related to a spin-polarized state ($S=3/2$) of three electrons. On the other
hand, a mixed-symmetry $\psi$ is related to a spin one half state.

Let us consider the lowest state with a given $J$, i.e. that one with
numbers $n_1=n_2=n_3=0$. This state may be spatially antisymmetrized only
when $J=3 k$, with $k$ - an integer. The argument goes along the lines
sketched in [9]: a cyclical permutation of electrons in the triangle,
leaving the wave function invariant, is equivalent to a $2\pi/3$ rotation,
which multiplies it by $e^{i J 2\pi/3}$, thus $e^{i J 2\pi/3}=1$. The
excitations with $n_2=n_3=0$ correspond also to antisymmetric states. The
$n_2+n_3=1$ states correspond to mixed-symmetry doublets, etc. On the other
hand, when $J\ne 3 k$ the lowest state and the excitations with $n_2=n_3=0$
have mixed symmetry. An antisymmetric and a mixed-symmetry state appear at
$n_2+n_3=1$, etc. We will restrict the analysis to the lowest state and the
first excitations.

The coefficient $a_2$ is thus given by

\begin{equation}
a_2 = \sqrt{3}(n_1+1/2) + \sqrt{3/2}(n_1+n_2+1).
\end{equation}

Higer corrections are computed from perturbation theory around $h_2$. By
the same reason as for two particles, $a_3$ and $a_5$ are equal to zero.
The next nonzero coefficient is

\begin{eqnarray}
a_4 &=&<n_1,n_2,n_3|h_4|n_1,n_2,n_3> \nonumber\\
    &+& \sum_{(k_1,k_2,k_3)\ne(n_1,n_2,n_3)}
      {<n_1,n_2,n_3|h_3|k_1,k_2,k_3><k_1,k_2,k_3|h_3|n_1,n_2,n_3>
       \over \sqrt{3}(n_1-k_1)+\sqrt{3/2}(n_1+n_2-k_1-k_2)}.
       \nonumber\\
\end{eqnarray}

The computation of matrix elements entering (35) is trivial, leading to

\begin{eqnarray}
a_4 &=& {1\over 144 \rho_{10}^2}
        \left\{-2+9\sqrt{2}+72 J^2 +(12+18\sqrt{2})n_1 +12 n_1^2
        \right. \nonumber\\
    &+& \left. (36+9\sqrt{2}+18\sqrt{2}n_1)(n_2+n_3)+228n_2 n_3-
        78(n_2^2+n_3^2) \right\}.
\end{eqnarray}

Let us stress that we expanded the hamiltonian around a structure with
$\theta_1=\pi/2$. There is an equivalent configuration with
$\theta_1=-\pi/2$. In the expansion (16), we have neglected tunneling
between the two equivalent configurations. The same comment holds for
systems with more than three electrons. Notice also that the second local
minimum of $U$, the linear structure (second ``Lagrange'' solution), is
at a distance 0.36$\beta^2$ above the lowest state. We can disregard any
effect comming from this structure when $\beta^2\gg a_2/0.36$.

The computation of the coefficient $b_3$ requires the wave functions at
$\beta=0$. They may be explicitly written with the help of the indications
of paper [10]. We will compute $b_3$ for a set of states with $|J|$=0, 1,
2, and 3. In all of these states, the corresponding quantum numbers at
$\beta\to\infty$ can be specified.

For example, when $J=0$, the first antisymmetric (A) state at $\beta=0$,
which starts from $b_0=4$ (it will be labelled (4,A)), goes to the first A
at $\beta\to\infty$, i.e. $(n_1,n_2,n_3)=(0,0,0)$. The first mixed state,
(4,M), goes to the first doublet with $n_1=0$, $n_2+n_3=1$.The second
antisymmetric state, (6,A), goes to the (1,0,0). At $|J|=1$, the (3,M) goes
to the (0,0,0), the (5,A) goes to a state with $n_1=0$, $n_2+n_3=1$. At
$|J|=2$, the (4,M) goes to the (0,0,0), and the (6,A) -- to a state with
$n_1=0$, $n_2+n_3=1$. At $|J|=3$, the (5,A) goes to the (0,0,0), and the
(5,M) to a state with $n_1=0$, $n_2+n_3=1$.

Let us write explicitly the needed wave functions at $\beta=0$ (up to
normalisations). In the case
of mixed symmetry, only one representative of the doublet is given.

$J=0$

\begin{eqnarray}
\phi_{4,A} &=& \rho_1 \rho_2 \sin \theta_1 \ e^{-(\rho_1^2+\rho_2^2)/4},\\
\phi_{4,M} &=& (\rho_1^2-\rho_2^2)\ e^{-(\rho_1^2+\rho_2^2)/4},\\
\phi_{6,A} &=& \rho_1 \rho_2 \sin \theta_1 \ (\rho_1^2+\rho_2^2-8)\
             e^{-(\rho_1^2+\rho_2^2)/4}.
\end{eqnarray}

$J=1$

\begin{eqnarray}
\phi_{3,M} &=& \rho_1\ e^{-i \theta_1/2}\ e^{-(\rho_1^2+\rho_2^2)/4},\\
\phi_{5,A} &=& \left\{(-2 \rho_1 \rho_2^2+\rho_1^3)\ e^{-i \theta_1/2}
             - \rho_1 \rho_2^2\ e^{3 i \theta_1/2} \right\}
             \ e^{-(\rho_1^2+\rho_2^2)/4}.
\end{eqnarray}

$J=2$

\begin{eqnarray}
\phi_{4,M} &=& \rho_1 \rho_2\ e^{-(\rho_1^2+\rho_2^2)/4},\\
\phi_{6,A} &=& \rho_1 \rho_2 \sin \theta_1 \
             \left(\rho_1^2 e^{-i \theta_1}+\rho_2^2 e^{i \theta_1}\right)\
             e^{-(\rho_1^2+\rho_2^2)/4}.
\end{eqnarray}

$J=3$

\begin{eqnarray}
\phi_{5,A} &=& \left\{\rho_1^3 e^{-3 i \theta_1/2}-
                 3\rho_1 \rho_2^2 e^{i \theta_1/2}\right\}\
            e^{-(\rho_1^2+\rho_2^2)/4},\\
\phi_{5,M} &=& \left\{\rho_1^3 e^{-3 i \theta_1/2}+
                 \rho_1\rho_2^2 e^{i \theta_1/2}\right\}\
                e^{-(\rho_1^2+\rho_2^2)/4}.
\end{eqnarray}

The computation of $b_3$ is thus a trivial task. We groupped all the results
together in Table 2. With these coefficients, we construct the approximants
$P_{3,2}$, $P_{4,3}$, and $P_{5,4}$.

The convergence analysis of the Pade sequence $P_{K+1,K}$ is shown in Fig. 4
for the lowest antisymmetric state with $|J|=3$. As in the two-electron
problem, convergence is strong, suggesting the $P_{5,4}$ interpolant to be
accurate to about 6 parts in $10^3$ or better in the entire interval
$0\le\beta<\infty$. Similar results are obtained for the lowest states with
$|J|=2$, and 1. The excited states and the states with $J=0$ show a slower
convergence. The relative error is estimated as 2\%.

\section{Four electrons}

First, let us compute the coefficients $a_0$ and $a_2$. As in previous
cases, $a_1=a_3=0$.

The equilibrium configuration at $\beta\to\infty$ is a square with side
$\rho_{10}=(2+\sqrt{2}/2)^{1/3}$ (scaled). The equilibrium values of the
other coordinates are $\rho_{20}=\sqrt{5/3}\ \rho_{10}$,
$\rho_{30}=2\rho_{10}/\sqrt{3}$, $\theta_{10}=\arctan{2}$,
$\theta_{20}=3\pi/4-\theta_{10}$. $a_0$ is given by

\begin{equation}
a_0 = 3(2+\sqrt{2}/2)^{1/3}.
\end{equation}

Expanding around the equilibrium geometry, we obtain the quadratic
hamiltonian,

\begin{eqnarray}
h_2 &=&-\left( {\partial^2 \over \partial z_1^2}+
               {\partial^2 \over \partial z_2^2}+
               {\partial^2 \over \partial z_3^2}\right)
       -\left( {1\over\rho_{10}^2}+{1\over\rho_{20}^2}\right)
               {\partial^2 \over \partial z_4^2}\nonumber\\
    &-&\left( {1\over\rho_{20}^2}+{1\over\rho_{30}^2}\right)
               {\partial^2 \over \partial z_5^2}
       +{2\over\rho_{20}^2}
               {\partial\over\partial z_4}{\partial\over\partial z_5} + V_2,
\end{eqnarray}

\noindent
where

\begin{eqnarray}
V_2 &=& 0.543374 z_1^2+0.157283 z_1 z_2+0.520305 z_2^2+0.182039 z_1 z_3
                         \nonumber\\
    &+& 0.377402 z_2 z_3+0.460201 z_3^2+0.171862 z_2 z_4+0.192148 z_3 z_4
                         \nonumber\\
    &+& 0.517308 z_4^2+0.110937 z_1 z_5 +0.085931 z_2 z_5+0.562904 z_4 z_5
                         \nonumber\\
    &+& 0.853613 z_5^2.
\end{eqnarray}

The normal frequencies are easily found, resulting in

\begin{eqnarray}
\omega_1 &=& 1.04969, \nonumber\\
\omega_2 &=& 1.13911, \nonumber\\
\omega_3 &=& 1.36227, \nonumber\\
\omega_4 &=& 1.46201, \nonumber\\
\omega_5 &=& 1.75640,
\end{eqnarray}

\noindent
and, thus, the coefficient $a_2$ is given by

\begin{equation}
a_2 = \sum_{k=1}^5 \omega_k (n_k+1/2).
\end{equation}

Notice that the frequency corresponding to the breathing mode, $\omega_5$,
is close to the classical value $\sqrt{3}$, but it does not exactly coincide
with $\sqrt{3}$.

As in the $N=3$ problem, the lowest state with a given $|J|$, i.e. $n_k=0$,
$k=1,\dots,5$, can be spatially antisymmetrised only when $|J|$ takes
certain values. The allowed values are $|J| = 2, 6, 10, \dots$. These are
polarised spin states, i.e. with total spin $S=2$. The excitations of the
$\omega_5$ mode have the symmetry of the ground state.

Let us stress that, as $N$ increases, the number of equivalent
configurations (geometries with the same $a_0$) increases, and the energetic
distance to other local minima of the classical energy decreases. Thus,
tunelling effects become more and more important.

In what follows, we restrict the analysis to the lowest spin-polarised
state, that is $|J|=2$, and the $n_k=0$. It seems that it is the lowest
state of $h$ in the sector with $S=2$ at any $\beta$. Indeed, as
$\beta\to 0$, this state goes to an oscillator state with $b_0=7$, i.e. to
the (7,A) in the terminology used above. There is a second (7,A) with
$J=0$. However, according to Hund's rule, this state has a higher energy
at $\beta\ll 1$. When $\beta\to\infty$, the $J=0$ state is also higher in
energy because $J=0$ is not compatible with $n_k=0$, thus excitation quanta
are needed.

Let us compute the coefficient $b_3$ for the $|J|=2$ state. The wave
function at $\beta=0$ is given in Appendix 2. Calculations may be carried
out analytically, leading to

\begin{equation}
b_3 = 6 {<\phi_{7,A}|\frac{1}{\rho_1}|\phi_{7,A}> \over
             <\phi_{7,A}|\phi_{7,A}>}
    = \frac{91}{64} \sqrt{2\pi}.
\end{equation}

Once computed $a_0$, $a_2$, $b_0$, and $b_3$, we may construct the
approximants $P_{3,2}$ and $P_{4,3}$, which are the main results of this
section. The relative error between $P_{3,2}$ and $P_{4,3}$ is shown in
Fig. 5, suggesting that $P_{4,3}$ may estimate the energy with an error not
greater than 2\% at intermediate values of $\beta$.

Pade approximants to other levels may be constructed in the same way.

\section{Five electrons}

We follow the same programme as in the $N=4$ problem, i.e. computation of
$a_0$, $a_2$, $b_0$, and $b_3$ and, from them, construction of the
approximants $P_{3,2}$ and $P_{4,3}$.

The equilibrium configuration at $\beta\to\infty$ is a pentagon with side
$\rho_{10}=1.30766$ (scaled). The equilibrium values of the
other coordinates are $\rho_{20}/\rho_{10}=1.44177$,
$\rho_{30}/\rho_{10}=1.53244$, $\rho_{40}/\rho_{10}=1.34500$,
$\theta_{10}=0.865925$ rad, $\theta_{20}=1.74428$ rad,
$\theta_{30}=-1.03941$ rad. $a_0$ is given by

\begin{equation}
a_0 = 9.28013~~.
\end{equation}

Expanding around the equilibrium configuration, we obtain the hamiltonian
$h_2$,

\begin{eqnarray}
h_2 &=&-\left( {\partial^2 \over \partial z_1^2}+
               {\partial^2 \over \partial z_2^2}+
               {\partial^2 \over \partial z_3^2}+
               {\partial^2 \over \partial z_4^2}\right)
       -\left( {1\over\rho_{10}^2}+{1\over\rho_{20}^2}\right)
               {\partial^2 \over \partial z_5^2}\nonumber\\
    &-&\left( {1\over\rho_{20}^2}+{1\over\rho_{30}^2}\right)
               {\partial^2 \over \partial z_6^2}
       -\left( {1\over\rho_{30}^2}+{1\over\rho_{40}^2}\right)
               {\partial^2 \over \partial z_7^2}\nonumber\\
   &+& {2\over\rho_{20}^2}
              {\partial\over\partial z_5}{\partial\over\partial z_6}
      + {2\over\rho_{30}^2}
              {\partial\over\partial z_6}{\partial\over\partial z_7} + V_2,
\end{eqnarray}

\noindent
where

\begin{eqnarray}
V_2 &=& 0.624037 z_1^2-0.007988 z_1 z_2+0.514889 z_2^2+0.107324 z_1 z_3
                         \nonumber\\
    &+& 0.231789 z_2 z_3+0.438330 z_3^2+0.073587 z_1 z_4+0.245901 z_2 z_4
                         \nonumber\\
    &+& 0.381949 z_3 z_4+0.373258 z_4^2+0.315672 z_2 z_5-0.296995 z_3 z_5
                         \nonumber\\
    &+& 0.451706 z_5^2+0.313498 z_1 z_6-0.217440 z_2 z_6-0.286484 z_3 z_6
                         \nonumber\\
    &+& 0.326409 z_4 z_6+0.765563 z_5 z_6+1.43839 z_6^2-0.126796 z_2 z_7
                         \nonumber\\
    &+& 0.119294 z_3 z_7+0.169244 z_5 z_7+1.37849 z_6 z_7+1.46079 z_7^2.
\end{eqnarray}

The normal frequencies following from the eigenvalue problem for $h_2$ are

\begin{eqnarray}
\omega_1 &=& 0.727516, \nonumber\\
\omega_2 &=& 0.804763, \nonumber\\
\omega_3 &=& 1.37185, \nonumber\\
\omega_4 &=& 1.52505, \nonumber\\
\omega_5 &=& 1.68899, \nonumber\\
\omega_6 &=& 1.73205, \nonumber\\
\omega_7 &=& 1.74375,
\end{eqnarray}

\noindent
and the coefficient $a_2$ is given by

\begin{equation}
a_2 = \sum_{k=1}^7 \omega_k (n_k+1/2).
\end{equation}

The allowed values of $|J|$ for antisymmetric states with $n_k=0$ are
$|J| = 0, 5, 10, \dots$. In these states the total spin is $S=5/2$. In what
follows, we consider the lowest state in this sequence, i.e. $J=0$. This
state goes to a (10,A) state as $\beta\to 0$. It is not, however, the
lowest of all antisymmetric levels at $\beta\ll 1$ because a second (10,A)
state with $|J|=2$ minimises the Coulomb repulsion.

The wave function of the $J=0$ state is given in Appendix 2. Calculations
may also be performed analytically, leading to

\begin{equation}
b_3 = \frac{149}{64} \sqrt{2\pi}.
\end{equation}

We show in Fig. 6 the relative difference between $P_{3,2}$ and $P_{4,3}$.
This difference is not greater than 2.5\%.

Pade approximants to other levels may be constructed in the same way.

\section{Concluding Remarks}

We have studied systems of 2 - 5 electrons in a two-dimensional parabolic
quantum dot. The potentials involved in this problem (harmonic plus
Coulomb repulsion) are very gentle, and lead to a smooth dependence of the
energy $\epsilon$ on the coupling constant $\beta$. This fact is graphically
represented in Fig. 1, where it is seen that the ``regions of convergence"
of the perturbative and the strong-coupling series ``intersect" with each
other.

The degrees of homogeneity of the potentials are also important factors
towards the smoothness of $\epsilon$. They lead to expansion series
containing only powers of $\beta^3$ at $\beta\to 0$, and inverse powers of
$\beta^2$ at $\beta\to\infty$. From the calculational point of view, it
means that with the help of trivial computations, not beyond first order
perturbation theory, we may construct approximants up to $P_{4,3}$, that is,
a quotient between a 5th order and a 3rd order polynomial in $\beta$. These
approximants are exact in both $\beta\to 0$ and $\beta\to\infty$ limits,
leading to errors not greater than 2.5 \% in the small transition region
where they jump from one expansion to the other. The accuracy may be
improved by computing higher approximants, as shown for two and three
electrons.

\section{Acknowledgments}

The author acknowledges financial support from the
Colombian Institute for Science and Technology (COLCIENCIAS) under Project
1118-05-661-95 and from the Committee for Scientific Research at the
Universidad Nacional (CINDEC). The author is endebted to B. Rodriguez, J.
Mahecha and L. Quiroga for useful discussions.
\newpage

\section{Appendix 1. The approximants used in the paper}

\begin{equation}
P_{3,2}(\beta) = b_0 + a_0 \beta^2 \left\{1-
                (1+b_3 \beta/a_0+a_0 \beta^2/(b_0-a_2))^{-1} \right\}.
\end{equation}

\noindent
$P_{4,3}$ is given in Eq. (7).

\begin{eqnarray}
P_{5,4}(\beta) &=& b_0 + {b_3 \beta^3 \over 1+q_1 \beta+\dots+q_4 \beta^4}
                 -{(b_0-a_2)q_4 \beta^4 \over 1+q_1 \beta+\dots+q_4 \beta^4}
                 \nonumber\\
               &+& a_0 \beta^2 \left\{1-{1+q_1\beta \over
                   1+q_1 \beta+\dots+q_4 \beta^4} \right\},
\end{eqnarray}

\noindent
where,

\begin{eqnarray}
q_2 &=& a_0 q_3/b_3,~~~q_1={1\over a_0}\{b_3+(b_0-a_2) q_3\},\nonumber\\
q_4 &=& {a_0 \over a_4} \left\{-1+{b_0-a_2 \over b_3} q_3 \right\},
\end{eqnarray}

\noindent
and $q_3$ is determined from the equation

\begin{equation}
-b_3 q_1 + a_0 q_2 -(b_0-a_2) q_4 = 0.
\end{equation}

\begin{eqnarray}
P_{6,5}(\beta) &=& b_0 + {b_3 \beta^3 \over 1+q_1 \beta+\dots+q_5 \beta^5}
                 -{(b_0-a_2)\{q_4 \beta^4 +q_5 \beta_5 \}
                  \over 1+q_1 \beta+\dots+q_5 \beta^5} \nonumber\\
               &+& a_0 \beta^2 \left\{1-{1+q_1\beta \over
                   1+q_1 \beta+\dots+q_5 \beta^5} \right\},
\end{eqnarray}

where,

\begin{eqnarray}
q_3 &=& {1\over b_3}\{a_0 q_4 -b_6\},~~~q_2={1\over b_0-a_2}\{a_0+a_4 q_4\},
              \nonumber\\
q_1 &=& {1 \over b_3} \{a_0 q_2-(b_0-a_2) q_4 \},\nonumber\\
q_5 &=& {1 \over a_4} \{b_3 +(b_0-a_2) q_3 - a_0 q_1 \},
\end{eqnarray}

\noindent
and $q_4$ is found from

\begin{equation}
-b_3 q_2 + a_0 q_3 -(b_0-a_2) q_5 = 0.
\end{equation}

\section{Appendix 2. Wave functions for 4 and 5 electrons at $\beta=0$}

We show the explicit form of the functions $\phi_{7,A}$, $N=4$, and
$\phi_{10,A}$, $N=5$. The $\vec r_j$ are measured from the c. m., and
$(\vec a \times \vec b)_z$ denotes the $z$ component (normal to the plane)
of the vectorial product.

\begin{eqnarray}
\phi_{7,A} &=& \left\{ \left [ \vec\rho_3 \cdot \vec r_1/\rho_3 +
              i (\vec\rho_3 \times \vec r_1)_z/\rho_3 \right ]^2
              \left [\vec r_2 \times \vec r_3 + \vec r_3 \times \vec r_4 +
                   \vec r_4 \times \vec r_2  \right ]_z \right. \nonumber\\
           &-& \left [ \vec\rho_3 \cdot \vec r_2/\rho_3 +
              i (\vec\rho_3 \times \vec r_2)_z/\rho_3 \right ]^2
              \left [\vec r_3 \times \vec r_4 + \vec r_4 \times \vec r_1 +
                     \vec r_1 \times \vec r_3  \right ]_z \nonumber\\
           &+& \left [ \vec\rho_3 \cdot \vec r_3/\rho_3 +
              i (\vec\rho_3 \times \vec r_3)_z/\rho_3 \right ]^2
              \left [\vec r_4 \times \vec r_1 + \vec r_1 \times \vec r_2 +
                     \vec r_2 \times \vec r_4  \right ]_z \nonumber\\
           &-& \left.
               r_4^2 \left [\vec r_1\times\vec r_2+\vec r_2\times\vec r_3 +
               \vec r_3\times\vec r_1  \right ]_z \right\}
                    e^{-(\rho_1^2+\rho_2^2+\rho_3^2)/4}
                    e^{2 i(\theta_1+2 \theta_2)/3}.
\end{eqnarray}

\begin{eqnarray}
\phi_{10,A} &=& e^{-(\rho_1^2+\rho_2^2+\rho_3^2+\rho_4^2)/4}
              \left\{ (\vec r_1 \cdot \vec r_2)(\vec r_1 \times \vec r_2)_z
                \left[\vec r_3\times\vec r_4+\vec r_4\times\vec r_5+
                      \vec r_5\times\vec r_3  \right]_z\right. \nonumber\\
            &+& (\vec r_2 \cdot \vec r_3)(\vec r_2 \times \vec r_3)_z
                \left[\vec r_4\times\vec r_5+\vec r_5\times\vec r_1+
                      \vec r_1\times\vec r_4  \right]_z\nonumber\\
            &+& (\vec r_3 \cdot \vec r_4)(\vec r_3 \times \vec r_4)_z
                \left[\vec r_5\times\vec r_1+\vec r_1\times\vec r_2+
                      \vec r_2\times\vec r_5  \right]_z\nonumber\\
            &+& (\vec r_4 \cdot \vec r_5)(\vec r_4 \times \vec r_5)_z
                \left[\vec r_1\times\vec r_2+\vec r_2\times\vec r_3+
                      \vec r_3\times\vec r_1  \right]_z\nonumber\\
            &+& (\vec r_5 \cdot \vec r_1)(\vec r_5 \times \vec r_1)_z
                \left[\vec r_2\times\vec r_3+\vec r_3\times\vec r_4+
                      \vec r_4\times\vec r_2  \right]_z\nonumber\\
            &-& (\vec r_1 \cdot \vec r_3)(\vec r_1 \times \vec r_3)_z
                \left[\vec r_4\times\vec r_5+\vec r_5\times\vec r_2+
                      \vec r_2\times\vec r_4  \right]_z\nonumber\\
            &-& (\vec r_2 \cdot \vec r_4)(\vec r_2 \times \vec r_4)_z
                \left[\vec r_5\times\vec r_1+\vec r_1\times\vec r_3+
                      \vec r_3\times\vec r_5  \right]_z\nonumber\\
            &-& (\vec r_3 \cdot \vec r_5)(\vec r_3 \times \vec r_5)_z
                \left[\vec r_1\times\vec r_2+\vec r_2\times\vec r_4+
                      \vec r_4\times\vec r_1  \right]_z\nonumber\\
            &+& (\vec r_1 \cdot \vec r_4)(\vec r_1 \times \vec r_4)_z
                \left[\vec r_5\times\vec r_2+\vec r_2\times\vec r_3+
                      \vec r_3\times\vec r_5  \right]_z\nonumber\\
            &+& \left. (\vec r_2 \cdot \vec r_5)(\vec r_2 \times \vec r_5)_z
                \left[\vec r_1\times\vec r_3+\vec r_3\times\vec r_4+
                      \vec r_4\times\vec r_1  \right]_z\right\}.
\end{eqnarray}

\newpage

{\Large Figure and Table Captions}
\vspace{1cm}

\begin{description}
\item{Tab. 1.} The first nonzero coefficients $b_k$ and $a_k$ for a set
               of two-electron states
\item{Tab. 2.} The first nonzero coefficients $b_k$ and $a_k$ for certain
               three-electron states
\item{Fig. 1.} Weak and strong-coupling expansions (dashed lines), and the
               $P_{6,5}$ approximant (solid line) for two electrons in a
               state with $|J|=3$, $n_1=0$
\item{Fig. 2.} Relative differences between consecutive approximants in the
               sequence $P_{K+1,K}$, with $K$ ranging from 2 to 5. The
               state with $|J|=3$, $n_1=0$ is shown
\item{Fig. 3.} Convergence of the Pade sequence $P_{K+1,K}$ at $\beta=
               (2 |J|+1)^{1/6}$. Two electrons in states with $n_1=0$ are
               considered. Solid lines represent the exact solutions found
               in [8]
\item{Fig. 4.} The same as in Fig. 2, now for the first antisymmetric state
               of three electrons with $|J|=3$. $K$ ranges from 2 to 4
\item{Fig. 5.} Relative difference between $P_{4,3}$ and $P_{3,2}$ for the
               lowest antisymmetric state of four electrons ($|J|=2$)
\item{Fig. 6.} The same as in Fig. 5, now for the lowest antisymmetric state
               of five electrons with angular momentum $J=0$
\end{description}
\newpage

\noindent
Table 1
\vspace{1cm}

\begin{tabular}{|l|l|l|l|l|l|l|}
\hline\hline
$|J|$, $n$ & $b_0$ & $b_3$ & $b_6$ & $a_0$ & $a_2$ & $a_4$\\
\hline\hline
0, 0 & 1 & 1.253314 & -0.345655 & 1.190551 & 0.866025 & -0.034998\\
0, 1 & 3 & 0.939986 & -0.087406 & 1.190551 & 2.598076 &  0.174989\\
0, 2 & 5 & 0.802904 & -0.041161 & 1.190551 & 4.330127 &  0.594963\\
\hline
1, 0 & 2 & 0.626657 & -0.032762 & 1.190551 & 0.866025 &  0.594963\\
1, 1 & 4 & 0.548325 & -0.016339 & 1.190551 & 2.598076 &  0.804950\\
1, 2 & 6 & 0.499367 & -0.010112 & 1.190551 & 4.330127 &  1.224924\\
\hline
2, 0 & 3 & 0.469993 & -0.011153 & 1.190551 & 0.866025 &  2.484844\\
2, 1 & 5 & 0.430827 & -0.007016 & 1.190551 & 2.598076 &  2.694831\\
2, 2 & 7 & 0.402676 & -0.004920 & 1.190551 & 4.330127 &  3.114805\\
\hline
3, 0 & 4 & 0.391661 & -0.005533 & 1.190551 & 0.866025 &  5.634647\\
3, 1 & 6 & 0.367182 & -0.003911 & 1.190551 & 2.598076 &  5.844634\\
3, 2 & 8 & 0.348211 & -0.002950 & 1.190551 & 4.330127 &  6.264608\\
\hline\hline
\end{tabular}

\vspace{2cm}

\noindent
Table 2
\vspace{1cm}

\begin{tabular}{|l|l|l|l|l|l|}
\hline\hline
$|J|$, Symmetry & $b_0$ & $b_3$ & $a_0$ & $a_2$ & $a_4$\\
\hline\hline
0, A & 4 & 1.87997 & 3.12013 & 2.09077 &  0.0358156\\
0, M & 4 & 2.58496 & 3.12013 & 3.31552 & -0.0619104\\
0, A & 6 & 1.76247 & 3.12013 & 3.82282 &  0.200926\\
\hline
1, M & 3 & 2.81996 & 3.12013 & 2.09077 &  0.276191\\
1, A & 5 & 1.82122 & 3.12013 & 3.31552 &  0.178464\\
\hline
2, M & 4 & 2.23247 & 3.12013 & 2.09077 &  0.997315\\
2, A & 6 & 1.58623 & 3.12013 & 3.31552 &  0.899589\\
\hline
3, A & 5 & 1.70372 & 3.12013 & 2.09077 &  2.19919\\
3, M & 5 & 2.26184 & 3.12013 & 3.31552 &  2.10146\\
\hline\hline
\end{tabular}

\end{document}